# Energy Efficient Virtual Machines Placement over Cloud-Fog Network Architecture

Hatem Alharbi, Taisir E.H. Elgorashi and Jaafar M.H. Elmirghani
School of Electronic and Electrical Engineering University of Leeds, LS2 9JT, United Kingdom

*Abstract*— Fog computing is an emerging paradigm that aims to improve the efficiency and QoS of cloud computing by extending the cloud to the edge of the network. This paper develops a comprehensive energy efficiency analysis framework based on mathematical modeling and heuristics to study the offloading of virtual machine (VM) services from the cloud to the fog. The analysis addresses the impact of different factors including the traffic between the VM and its users, the VM workload, the workload versus number of users profile and the proximity of fog nodes to users. Overall, the power consumption can be reduced if the VM users' traffic is high and/or the VMs have a linear power profile. In such a linear profile case, the creation of multiple VM replicas does not increase the computing power consumption significantly (there may be a slight increase due to idle / baseline power consumption) if the number of users remains constant, however the VM replicas can be brought closer to the end users, thus reducing the transport network power consumption. In our scenario, the optimum placement of VMs over a cloud-fog architecture significantly decreased the total power consumption by 56% and 64% under high user data rates compared to optimized distributed clouds placement and placement in the existing AT&T network cloud locations, respectively.

*Index Terms*— **Fog computing, IP over WDM network, energy efficiency, virtual machine, workload profile**

## I. INTRODUCTION

Cloud Computing has started to transform the information and communication technology (ICT) industry by providing efficient resource sharing solutions where an Internet-based pool of network, storage and computational resources is made available to simultaneously serve a large number of geographically distributed users. Cloud computing is an essential enabler for the development of emerging IoT and Big Data applications. By 2020, the total cloud traffic is expected to grow to 3.7 times its level in 2015 reaching 1.2 zetta bytes per month which accounts for 92% of the total data center traffic [1]. This mounting traffic creates a huge burden on data centers and networks leading to serious challenges in terms of energy efficiency and QoS [2].

The concept of fog computing, introduced by Cisco in 2014 [3], came to complement the central cloud services by offloading some of the services into the geographical proximity of users in the network edge so that services can be efficiently accessed. The research efforts in fog computing have mainly focused on illustrating its potential advantages over cloud computing. Fog computing is proposed to provide low latency [4], conserve network bandwidth [5], improve quality-of-services (QoS) [6] and quality-of-experience (QoE) [7] for different computing services.

The energy consumption of cloud and fog computing has been given limited attention in the literature. In [8], the authors found that the number of hops between the user and the content has little impact on the total energy consumption compared to the type of application running on servers and factors such as the number of downloads and number of updates. In [9], the authors studied the interplay and cooperation between the fog and the cloud to achieve a tradeoff between power consumption and delay in a cloud-fog computing system. A detailed analysis of the essential service metrics with regard to the cost and benefit of offloading services to the fog layer is yet to be conducted to identify the services that the fog can efficiently host. Such an analysis is crucial to sustain the growth of the IoT and Big Data applications which are proving to be pivotal to economic growth and quality of life. In [10], the authors built a theoretical model of fog computing architecture and compared it with the conventional cloud computing model. In addition to the low latency, they found that offloading applications to the fog layer can significantly reduce power consumption by 41%. However, their investigation did not consider a detailed model of the telecom network architecture. The authors of [11], [12] developed models for the core network energy efficiency, while [13] focused on developing energy efficient network topologies, with the core optical network energy efficiency being considered in [14] – [16]. The carbon emissions of the network can further be reduced by introducing renewable energy into the network and optimizing its use, which is evaluated in [17]. The resilience of the network and its impact on energy efficiency was assessed in [18], [19], while energy efficient networking for content distribution was introduced in [20] – [23]. The optimization of the network for energy efficient big data transport was reported in [24], [25] while the use of analytics to optimize such networks was evaluated in [26], [27]. Finally, energy efficient network virtualization was evaluated in [28] – [30]. These studies did not however consider the interplay of cloud and fog processing and the optimum placement of virtual machines in the network to minimize the overall network and computing power consumption.

In this paper, we develop a comprehensive framework based on mathematical modeling and heuristics to study the offloading of virtual machine (VM) services from the cloud to the fog layer taking into consideration minimizing the total power consumption of providing the service. We optimize the placement of VMs over an end-to-end cloud-fog architecture that traverses the core network, metropolitan (metro) network and access network. The placement of VMs in the cloud at the core network allows VMs to serve users distributed across the core nodes whereas placing the VM replicas closer to the users in the fog nodes in the metro or access network limits the traffic between users and VMs to the metro and access networks respectively, thus eliminating the associated core network traffic (and potentially the metro traffic). This therefore reduces the network power consumption, but increases the processing power consumption due to the creation of multiple replicas of the VMs, and therefore a trade-off exists.

The remainder of this paper is organized as follows. Section II discusses the concept of machine virtualization and VMs workload profile and introduces the MILP model for optimizing the VM placement in the cloud-fog architecture. We present the optimization model results and analyze them in Section III. A real-time VM placement heuristic is proposed in Section IV. Finally, Section V concludes the paper.

## II. ENERGY EFFICIENT PLACEMENT OF VMS OVER CLOUD-FOG ARCHITECTURE

### A. *Machine Virtualization*

Cloud and fog processing employ Virtual Machines (VMs) for efficient resource utilization. Virtualization abstracts the server resources including the CPU, RAM, hard disk and I/O network to create an isolated virtual entity that can run its operating system and applications. The existence of such a virtual environment allows the scaling up and down of server resources in a dynamic manner based on the variation in user demands [31]. Further dynamism can be achieved by migrating or replicating VMs over geo-distributed servers to achieve different features such as load balancing [32] and energy efficiency [33]. The problem of migration and replication of VMs is referred to as VMs placement. VMs placement needs to be optimized to follow variations in the VMs demands, workload of the cloud/fog resources or network status [34].

In the literature, several papers discussed the VMs placement considering various factors. To reduce the server load, improve the QoS and meet the SLAs, the VMs can be migrated or replicated to another server/servers within the same datacentre [35] or in geographically distributed datacenters [36]. Virtualized cloud architectures are also able to provide efficient disaster resilience in case of physical machine failure by migrating VMs into different host machines [37] or by replicating VMs content to distributed datacenters. From an energy efficiency perspective, under-utilized servers can significantly increase the energy consumption, and consequently increase the carbon emissions and operating costs of cloud datacenters. VMs consolidation by bin packing them into a fewer number of servers can significantly improve the energy efficiency. The majority of studies of VMs placement in the fog have been limited to evaluating the reduction of overall network overhead [38], optimizing the placement of physical resources in the edge network [39] and the scheduling of VMs to share the limited fog resources to minimize SLA violations [40].

Despite the diverse factors affecting the power consumption of cloud-fog architectures, the problem of providing energy-efficient VMs placement over end-to-end cloud-fog architecture considering end-to-end architecture has not received any attention. Thus, the objective of this paper is to develop a novel framework that covers different networks and computing in optimizing the energy efficiency of VMs placement.

### B. *VM workload profile:*

The power consumption of a VM is determined by its hosting server. The authors in [41] found that the CPU utilization and power consumption of a server are highly correlated. Another work in [42] studied the relationship between the power consumption of a server and the CPU utilization and found that the power consumption and the CPU utilization are related linearly. Thus, the work introduced in this thesis follows the same approach and takes into consideration the CPU utilization only in modelling the power consumption of VMs placement.

From a CPU perspective, studies in the literature have shown that the workload of VM versus the number of users served by VM mostly follows one of two profiles; constant profile or linear profile as seen in Fig. 1. In [43], the authors presented a CPU performance benchmark study for web application VMs serving a varying number of users with constant CPU workload as illustrated in Fig. 1 (a). Also, various benchmarking studies in the literature have demonstrated linear workload profiles for database applications [44], web-based video conferencing systems [45] and multiplayer games [46] with different slope coefficients. To maintain the service level agreement (SLA), each VM needs a minimum workload to run an application regardless of the number of users served by the VM, resulting in the workload profile shown in Fig. 1 (b). The minimum workload required to serve a user in a VM varies from as low as 1% to 60% [44] - [46].

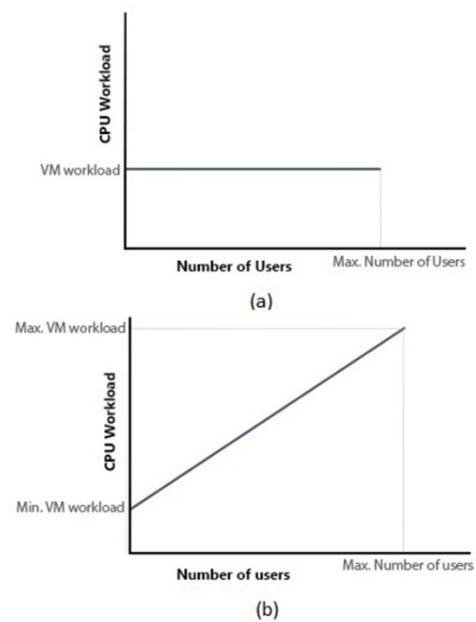

Figure 1: Relationship between VM workload and the number of users; (a) constant (b) linear relationship between VM workload and number of users.

### C. *MILP Model:*

In this section, we introduce the MILP model developed to optimize the placement of VMs over the cloud-fog architecture so that the power consumption of providing the VM services is minimized. We consider the architecture in Fig. 2 where a cloud layer is introduced at the core network and two fog layers are introduced at the metro network and the access network. In this work, we take a general approach in the placement of VMs over the cloud-fog architecture unlike [33] where certain placement schemes are imposed on all types of VMs. We allow the MILP model to select the most energy efficient placement for each VM based on VM popularity, the VM minimum workload requirement and data rate. The model aims to achieve a trade-off between network power saved by replicating VMs in multiple clouds and/or fog nodes and the power consumed by these replicas. The creation of a VM replica results in power savings if the former power exceeds the latter power.

Before introducing the model, we define the parameters and variables related to the different layers of the cloud-fog architecture in Fig.2.

1) *Cloud and fog nodes*

A typical data center, as illustrated in Fig. 2, consists of servers arranged in multiple racks and a LAN network, made of routers and switches, to connect racks to each other (inter rack communication) and to users outside the data center. The resources at the fog nodes form mini data centers connected in a similar way to the cloud data centers. Servers, switches and routers in the cloud and fog nodes are defined by the following parameters:

Cloud and fog parameters

| | |
|---|---|
| $SW^{(CB)}$ | Cloud switch bit rate. |
| $SW^{(CP)}$ | Cloud switch power consumption. |
| $SW^{(MFB)}$ | Metro fog switch bit rate. |
| $SW^{(MFP)}$ | Metro fog switch power consumption. |
| $SW^{(AFB)}$ | Access fog switch bit rate. |
| $SW^{(AFP)}$ | Access fog switch power consumption. |
| $SW^{(R)}$ | Cloud and fog switch redundancy. |
| $R^{(CB)}$ | Cloud router port bit rate. |
| $R^{(CP)}$ | Cloud router port power consumption. |
| $R^{(MFB)}$ | Metro fog router port bit rate. |
| $R^{(MFP)}$ | Metro fog router port power consumption. |
| $R^{(AFB)}$ | Access fog router port bit rate. |
| $R^{(AFP)}$ | Access fog router port power consumption. |
| $S^{(P)}$ | Power consumption of a server. |
| $S^{(maxW)}$ | Maximum workload of a server. |
| $c$ | Cloud power usage effectiveness. |
| $m$ | Metro fog power usage effectiveness. |
| $a$ | Access fog power usage effectiveness. |

Cloud and fog variables

| | |
|---|---|
| $C_s$ | $C_s = 1$ if a cloud is hosted in node $s$, otherwise $C_s = 0$. |
| $\delta_{v,s}^{(C)}$ | $\delta_{v,s}^{(C)} = 1$ if the cloud hosted in node $s$ hosts a copy of VM $v$, otherwise $\delta_{v,s}^{(C)} = 0$. |
| $R_s^{(C)}$ | Number of router aggregation ports in the cloud hosted in node $s$. |
| $SW_s^{(C)}$ | Number of switches in the cloud hosted in node $s$. |
| $S_s^{(C)}$ | Number of processing servers in the cloud hosted in node $s$. |
| $F_s^{(MF)}$ | $F_s^{(MF)} = 1$ if a fog processing node is hosted in the metro network connected to core node $s$, otherwise $F_s^{(MF)} = 0$. |
| $\delta_{v,s}^{(MF)}$ | $\delta_{v,s}^{(MF)} = 1$ if the fog processing node hosted in the metro network connected to node $s$ hosts a replica of VM $v$, otherwise $\delta_{v,s}^{(MF)} = 0$. |
| $R_s^{(MF)}$ | Number of router ports used in the fog processing node hosted in the metro network connected to node $s$. |
| $SW_s^{(MF)}$ | Number of switches used in the fog processing node hosted in the metro network connected to node $s$. |
| $S_s^{(MF)}$ | Number of processing servers in the fog processing node hosted in the metro network connected to node $s$. |
| $F_{p,s}^{(AF)}$ | $F_{p,s}^{(AF)} = 1$ if a fog processing node is built in access network $p$ connected to core node $s$, otherwise $F_{p,s}^{(AF)} = 0$. |
| $\delta_{v,p,s}^{(AF)}$ | $\delta_{v,p,s}^{(AF)} = 1$ if the fog processing node in access network $p$ connected to core node $s$, hosts a replica of VM $v$, otherwise $\delta_{v,p,s}^{(AF)} = 0$. |
| $R_{p,s}^{(AF)}$ | Number of router ports used in the fog processing node located in the access network $p$ connected to core node $s$. |
| $SW_{p,s}^{(AF)}$ | Number of switches used in the fog processing node located in access network $p$ connected to core node $s$. |
| $S_{p,s}^{(AF)}$ | Number of processing servers in the fog processing node located in the access network $p$ connected to core node $s$. |

The VMs to be hosted in the cloud and/or fog and the traffic resulting from them are defined by the following parameters and variables:

VM parameters

| | |
|---|---|
| $N$ | Set of IP over WDM network nodes. |
| $VM$ | Set of VM services. |
| $s$ and $d$ | Indices of source and destination nodes of a traffic flow in the distributed cloud architecture. |
| $V$ | Number of VMs. |
| $S_v$ | Number of VM $v$ users. |
| $r_v$ | User download rate of VM $v$. |

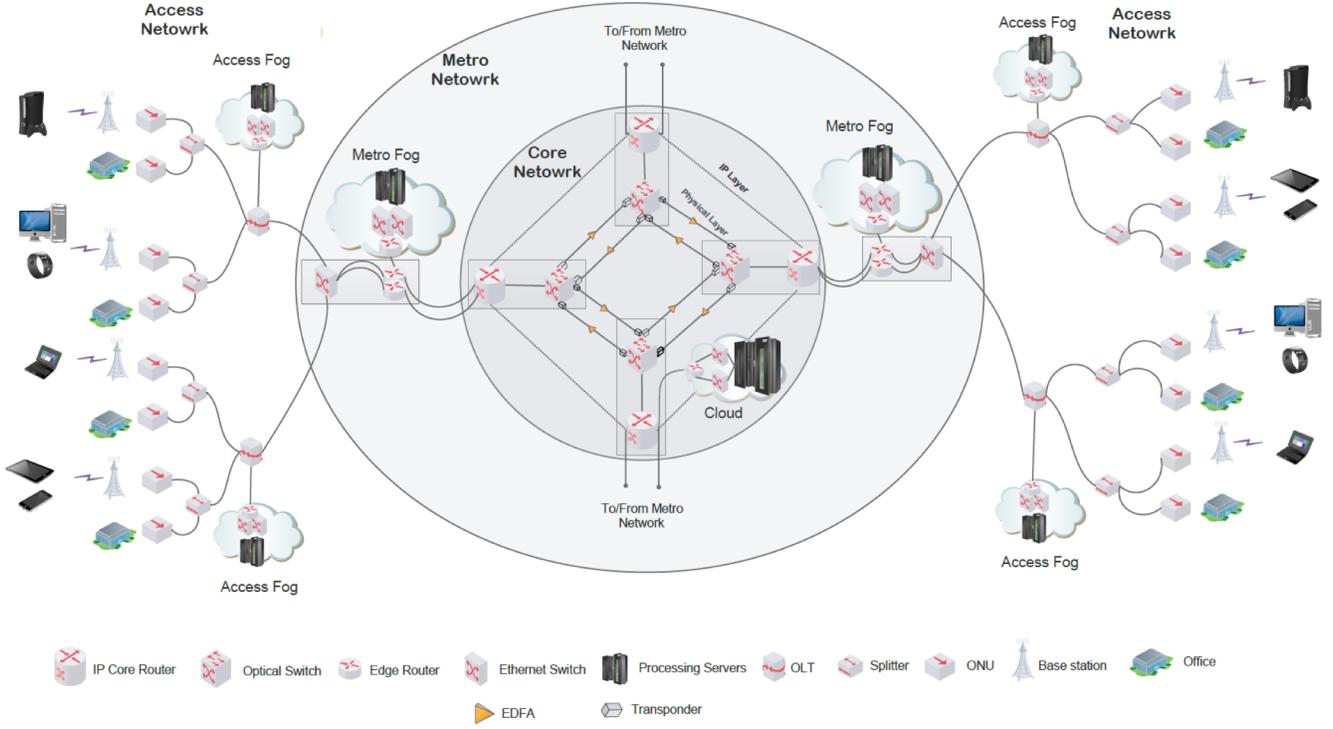

Figure 2: Cloud-Fog architecture

| | |
|---|---|
| $L$ | Large enough number. |
| $x$ | Maximum number of users served by a single VM replica. |
| $W_v$ | Maximum workload of VM $v$ (workload can be specified in GHz or as a ratio of the CPU capacity). |
| $M$ | Workload baseline of VM (the minimum CPU utilization needed in the absence of load). |
| $T_v$ | Traffic resulting from VM replica $v$ serving the maximum number of users. |

$$T_v = x\, r_v$$

| | |
|---|---|
| $W_v^{(R)}$ | Workload per traffic unit, |

$$W_v^{(R)} = \frac{W_v - M}{T_v}$$

evaluated for VM replica $v$.

VM variables

| | |
|---|---|
| $W_{v,s}^{(CR)}$ | Workload of VM replica $v$ hosted in cloud in node $s$. |
| $W_s^{(C)}$ | Total workload of cloud hosted in node $s$. |
| $D_{v,s,d}^{(C)}$ | Traffic flow from VM replica $v$ hosted in cloud of node $s$ to users in node $d$. |
| $L_{s,d}$ | Traffic from cloud node $s$ to users in node $d$. |
| $W_{v,s}^{(MFR)}$ | Workload of the VM replica $v$ hosted in the fog processing node located in the metro network connected to node $s$. |
| $W_s^{(MF)}$ | Total workload of the metro fog processing node located in core node $s$. |
| $D_{v,s}^{(MF)}$ | Traffic from the VM replica $v$ hosted in the fog processing node of the metro network connected to core node $s$. |
| $W_{v,p,s}^{(AFR)}$ | Workload of the VM replica $v$ hosted in the fog processing node located in the access network $p$ connected to core node $s$. |
| $W_{p,s}^{(AF)}$ | Total workload of the fog processing node located in the access network $p$ connected to core node $s$. |
| $D_{v,p,s}^{(AF)}$ | Traffic flow from the VM replica $v$ hosted in the fog processing node located in the access network $p$ connected to core node $s$. |

The clouds power consumption ($CLOUD$) is composed of:

1) Power consumption of cloud servers:

$$c \sum_{s \in N} S_s^{(C)} S^{(P)} \qquad (1)$$

2) Power consumption of cloud routers and switches:

$$c \left( \sum_{s \in N} \left( (SW_s^{(C)} SW^{(R)} SW^{(CP)}) + R_s^{(C)} R^{(CP)} \right) \right) \qquad (2)$$

The metro fogs ($MF$) power consumption is composed of:

1) Power consumption of metro fog servers:

$$m \sum_{s \in N} S_s^{(MF)} S^{(P)} \qquad (3)$$

2) Power consumption of metro fog switches and routers:

$$m\left(\sum_{s\in N}\left(\left(SW_s^{(MF)} SW^{(R)} SW^{(MFP)}\right) + R_s^{(MF)} R^{(MFP)}\right)\right) \quad (4)$$

The access fogs power consumption ($AF$) is composed of:

1) Power consumption of access fog servers:

$$a\sum_{s\in N}\sum_{p\in P} S_{p,s}^{(AF)} S^{(P)} \quad (5)$$

2) Power consumption of access fog switches and routers:

$$a\left(\sum_{s\in N}\sum_{p\in P}\left(\left(SW_{p,s}^{(AF)} SW^{(R)} SW^{(AFP)}\right) + R_{p,s}^{(AF)} R^{(AFP)}\right)\right) \quad (6)$$

Note that, as the difference between the server idle power and full load is very small [47], we consider an on-off power profile for servers, i.e. if a server is activated, it operates at maximum power consumption.

2) *Access network*

Passive optical networks (PONs) [48] are the selected technology for the access network in the cloud-fog architecture given in Fig. 1 due to their high bandwidth, reliability, and high data transmission compared to Ethernet access networks. At present, the gigabit PON (GPON) architecture has become the most popular solution for PON among service providers [49]. Two main active components are deployed in GPON; the optical network unit (ONU) and the optical line terminal (OLT). The ONU is the end-user interface to the PON network and the OLT serves as a central office (CO) node to connect multiple ONUs. The Optical distribution networking [50] provides a passive physical transmission between OLT and ONU. XGPON is capable of delivering data rate up to 10 Gbps over a single port. In this work, we consider 10G-PON as an example of the PON network.

The following parameters and variables are defined to represent PON networks:

Access network parameters:

$P$      Set of PON networks.

$A_p$      Average broadband data rate in PON $p$.

$\Phi_v$      Ratio of traffic due to VM $v$ to the total PON traffic.

$OLT_{p,d}^{(B)}$      Capacity of OLT serving PON $p$ connected to node $d$.

$U_{v,p,d}$      Number of users in PON $p$ connected to core node $d$ requesting VM $v$.

$$U_{v,p,d} = \left(\frac{OLT_{p,d}^{(B)}}{A_p}\right)\Phi_v$$

if typical national/regional values of $A_p$, $\Phi_v$ and $OLT_{p,d}^{(B)}$ are used, then $U_{v,p,d}$ determines the number of users and their VM popularity.

$OLT_{p,d}^{(N)}$      Number of OLTs in PON network $p$ connected to node $d$.

$OLT^{(P)}$      OLT power consumption.

$D_{v,p,d}$      Traffic flow from VM $v$ to users in PON network $p$ connected to core node $d$ given as:

$$D_{v,p,d} = U_{v,p,d}\, r_v$$

$ONU_{p,d}^{(N)}$      Number of ONUs in PON network $p$ connected to node $d$.

$ONU^{(P)}$      Power consumption of an ONU.

$n$      Network power usage effectiveness.

PON networks power consumption ($PON$) is composed of:

1) Total power consumption of OLT:

$$n\left(\sum_{p\in P}\sum_{d\in N}\left(OLT^{(P)} OLT_{p,d}^{(N)}\right)\right) \quad (7)$$

2) Total power consumption of ONUs:

$$n\left(\sum_{p\in P}\sum_{d\in N}\left(ONU^{(P)} ONU_{p,d}^{(N)}\right)\right) \quad (8)$$

3) *Metro network*

A metro network [51] functions as a gateway for the access networks into the core network. Metro Ethernet is the dominant technology used in enterprise metro network. The basic components of metro Ethernet are Ethernet switch and edge routers as shown in Fig. 2. The Ethernet switch interconnects several access networks together. Also, it connects the access networks to edge routers. The best practice in ISP metro network is to use two edge routers in order to provide reliability and redundancy to the network [52]. The following parameters are defined to represent the metro network.

Metro network parameters:

$R^{(MB)}$      Metro router bit rate.

$R^{(MP)}$      Metro router power consumption.

$R^{(MR)}$      Metro router redundancy.

$SW^{(MB)}$      Metro Ethernet switch bit rate.

$SW^{(MP)}$      Metro Ethernet power consumption.

Metro network variables:

$R_s^{(M)}$      Number of router ports in metro network connected to node $s$.

$SW_s^{(M)}$      Number of Ethernet switches in metro network connected to node $s$.

The metro network power consumption ($Metro$) is composed of:

1) Total power consumption of edge routers:

$$n\left(\sum_{s\in N} R_s^{(M)} R^{(MR)} R^{(MP)}\right) \quad (9)$$

2) Total power consumption of edge Ethernet switches:

$$n\left(\sum_{s\in N} SW_s^{(M)} SW^{(MP)}\right) \quad (10)$$

## 4) Core network

The IP over WDM network [53] is the most commonly used architecture in core networks. The components of the IP layer and physical layer are shown on Fig. 2. In the IP layer, the core router controls the Internet traffic. It aggregates the IP traffic packets from the edge router to be sent to their destination. Optical switches make the connection between physical layer and IP layer. Optical switches are connected to fiber links. In each switching node, the transponder provides optical-electronic-optical (OEO) conversion for full wavelength conversion. In addition, for long distance transmission, erbium-doped fiber amplifiers (EDFAs) are used to amplify the optical signal in each fiber [53]. Regenerators are used to re-amplify, re-shape and re-time (3R) the optical signal in long-haul transmission [54]. The IP over WDM network can be implemented using either the non-bypass approach or the lightpath bypass approach. Under the non-bypass approach, the packets are processed by the IP layer of every intermediate node during their journey from the source to destination. On the other hand, under the bypass approach, the intermediate nodes introduce a shortcut by bypassing the IP layer (of intermediate nodes) on the way to the destination node.

The following parameters and variables are defined to represent the IP over WDM core network:

Core network parameters:
| | |
|---|---|
| $m$ and $n$ | Indices of the end nodes of a physical link. |
| $i$ and $j$ | Indices of the end nodes of a virtual link. |
| $Nm_m$ | Set of neighbouring nodes of node $m$. |
| $R^{(P)}$ | Core router port power consumption. |
| $t^{(P)}$ | Transponder power consumption. |
| $e^{(P)}$ | EDFA power consumption. |
| $SW_s^{(P)}$ | Optical switch power consumption in node $s$. |
| $G^{(P)}$ | Regenerator power consumption. |
| $\mathcal{W}$ | Number of wavelengths per fibre. |
| $\mathcal{W}^{(B)}$ | Wavelength data rate. |
| $S$ | Maximum span distance between two EDFAs in kilometres. |
| $D_{m,n}$ | Distance in kilometres between node pair $(m,n)$. |
| $A_{m,n}$ | Number of EDFAs between node pair $(m,n)$. $A_{m,n} = \left\lfloor \frac{D_{m,n}}{S} - 1 \right\rfloor$ where $S$ is the reach of the EDFA. |
| $G_{m,n}$ | Number of regenerators between node pair $(m,n)$. Typically $G_{m,n} = \left\lfloor \frac{D_{m,n}}{R} - 1 \right\rfloor$, where $R$ is the reach of the regenerator. |

Core network variables:
| | |
|---|---|
| $C_{i,j}$ | Number of wavelengths in virtual link $(i,j)$. |
| $\mathcal{W}_{m,n}$ | Number of wavelengths in physical link $(m,n)$. |
| $R_s^{(AC)}$ | Number of router ports in node $s$ that aggregate the traffic from/to clouds. |
| $R_d^{(AE)}$ | Number of router ports in node $d$ that aggregate the traffic from/to metro routers. |
| $F_{m,n}$ | Number of fibres on physical link $(m,n)$. |
| $L_{i,j}^{s,d}$ | Amount of traffic flow between node pair $(s,d)$ traversing virtual link $(i,j)$. |
| $\mathcal{W}_{m,n}^{i,j}$ | Number of wavelengths of virtual link $(i,j)$ traversing physical link $(m,n)$. |

Under the non-bypass approach, the IP over WDM network power consumption ($Core$) is composed of [53]:

1. The power consumption of router ports:

$$n \left( \sum_{s \in N} R^{(P)} R_s^{(AC)} + \sum_{d \in N} R^{(P)} R_d^{(AE)} \right.$$
$$\left. + \sum_{m \in N} \sum_{n \in Nm_m : n \neq m} R^{(P)} \mathcal{W}_{m,n} \right) \quad (11)$$

2. The power consumption of transponders:

$$n \left( \sum_{m \in N} \sum_{n \in Nm_m : n \neq m} t^{(P)} \mathcal{W}_{m,n} \right) \quad (12)$$

3. The power consumption of EDFAs:

$$n \left( \sum_{m \in N} \sum_{n \in Nm_m : n \neq m} e^{(P)} F_{m,n} A_{m,n} \right) \quad (13)$$

4. The power consumption of optical switches:

$$n \left( \sum_{s \in N} SW_s^{(P)} \right) \quad (14)$$

5. The power consumption of regenerator:

$$n \left( \sum_{m \in N} \sum_{n \in Nm_m : n \neq m} G^{(P)} G_{m,n} \mathcal{W}_{m,n} \right) \quad (15)$$

The model is defined as follows:

The objective: *Minimize total power consumption given as the sum of the power consumptions:*

$$Core + Metro + PON + CLOUD + MF + AF \quad (16)$$

Expression (16) gives the total power consumption as the sum of the power consumption of the IP over WDM core network, the metro network, the PON access network, clouds, metro fogs and access fogs.

Subject to:

Serving VM demand constraints:

$$\sum_{p \in P} \sum_{d \in N} D_{v,p,d} = \sum_{s \in N} \sum_{d \in N} D_{v,s,d}^{(C)} + \sum_{s \in N} D_{v,s}^{(MF)} + \sum_{p \in P} \sum_{s \in N} D_{v,p,s}^{(AF)}$$
$$\forall \ v \in VM \quad (17)$$

Constraint (17) ensures that all the users demands for a VM are served by the clouds, the metro fogs or the access fogs.

Placing VM in cloud constraints:

$$L \sum_{d \in N} D_{v,s,d}^{(C)} \geq \delta_{v,s}^{(C)}$$
$$\forall \, s \in N, v \in VM \tag{18}$$

$$\sum_{d \in N} D_{v,s,d}^{(C)} \leq L \, \delta_{v,s}^{(C)}$$
$$\forall \, s \in N, v \in VM \tag{19}$$

Constraints (18) and (19) relate the binary variable that indicates whether a VM is hosted in a cloud or not ($\delta_{v,s}^{(C)}$) to the traffic between users of this VM and the cloud ($\sum_{d \in N} D_{v,s,d}^{(C)}$) by setting $\delta_{v,s}^{(C)} = 1$ if $\sum_{d \in N} D_{v,s,d}^{(C)} > 0$ and $\delta_{v,s}^{(C)} = 0$ otherwise.

Placing VM in metro fog constraints:

$$D_{v,s}^{(MF)} \geq \delta_{v,s}^{(MF)}$$
$$\forall \, s \in N, v \in VM \tag{20}$$

$$D_{v,s}^{(MF)} \leq L \, \delta_{v,s}^{(MF)}$$
$$\forall \, s \in N, v \in VM \tag{21}$$

Constraints (20) and (21) relate the binary variable that indicates whether a VM is hosted in a fog or not ($\delta_{v,s}^{(MF)}$) to the traffic between users of this VM and the metro fog ($D_{v,s}^{(MF)}$) by setting $\delta_{v,s}^{(MF)} = 1$ if $D_{v,s}^{(MF)} > 0$ and $\delta_{v,s}^{(MF)} = 0$ otherwise.

Placing VM in access fog constraints:

$$D_{v,p,s}^{(AF)} \geq \delta_{v,p,s}^{(AF)}$$
$$\forall \, s \in N, v \in VM, p \in P \tag{22}$$

$$D_{v,p,s}^{(AF)} \leq L \, \delta_{v,s}^{(AF)}$$
$$\forall \, s \in N, v \in VM, p \in P \tag{23}$$

Constraints (22) and (23) relate the binary variable that indicates whether a VM is hosted in an access fog or not ($AF\delta_{vsp}$) to the traffic between users of this VM and the cloud ($D_{v,p,s}^{(AF)}$), by setting $AF\delta_{vsp} = 1$ if $D_{v,p,s}^{(AF)} > 0$ and $\delta_{v,s}^{(AF)} = 0$ otherwise.

Clouds locations constraints:

$$\sum_{v \in VM} \delta_{v,s}^{(C)} \geq C_s$$
$$\forall \, s \in N \tag{24}$$

$$\sum_{v \in VM} \delta_{v,s}^{(C)} \leq L \, C_s$$
$$\forall \, s \in N \tag{25}$$

Constraints (24) and (25) ensure that a cloud is built in core nodes selected to host VMs by setting $C_s = 1$ if $\sum_{v \in VM} \delta_{v,s}^{(C)} > 0$ and $C_s = 0$ otherwise.

Metro fogs location constraints:

$$\sum_{v \in VM} \delta_{v,s}^{(MF)} \geq F_s^{(MF)}$$
$$\forall \, s \in N \tag{26}$$

$$\sum_{v \in VM} \delta_{v,s}^{(MF)} \leq L \, F_s^{(MF)}$$
$$\forall \, s \in N \tag{27}$$

Constraints (26) and (27) ensure that metro fogs are built in metro nodes selected to host VMs are by setting $MFog_s = 1$ if $\sum_{v \in VM} \delta_{v,s}^{(MF)} > 0$ and $MFog_s = 0$ otherwise.

Access fog location constraints:

$$\sum_{v \in VM} \delta_{v,p,s}^{(AF)} \geq F_{p,s}^{(AF)}$$
$$\forall \, s \in N \tag{28}$$

$$\sum_{v \in VM} \delta_{v,p,s}^{(AF)} \leq L \, F_{p,s}^{(AF)}$$
$$\forall \, s \in N \tag{29}$$

Constraints (28) and (29) ensure that an access fog is built in access nodes selected to host VMs by setting $F_{p,s}^{(AF)} = 1$ if $\sum_{v \in VM} \delta_{v,p,s}^{(AF)} > 0$ and $F_{p,s}^{(AF)} = 0$ otherwise.

Cloud and fog workload constraints:

$$W_{v,s}^{(CR)} = \delta_{v,s}^{(C)} \, W_v$$
$$\forall \, v \in VM, s \in N \tag{30}$$

$$W_{v,s}^{(CR)} = \left( \frac{\sum_{d \in N} D_{v,s,d}^{(C)}}{r_v \, x} \, M \, \delta_{v,s}^{(C)} \right) + \left( W_v^{(R)} \sum_{d \in N} D_{v,s,d}^{(C)} \right)$$
$$\forall \, v \in VM, s \in N \tag{31}$$

$$W_s^{(C)} = \sum_{v \in VM} W_{v,s}^{(CR)}$$
$$\forall \, s \in N \tag{32}$$

$$W_{v,s}^{(MFR)} = \delta_{v,s}^{(MF)} \, W_v$$
$$\forall \, v \in VM, s \in N \tag{33}$$

$$W_{v,s}^{(MFR)} = \left( \frac{D_{v,s}^{(MF)}}{r_v \, x} \, M \, \delta_{v,s}^{(MF)} \right) + \left( W_v^{(R)} D_{v,s}^{(MF)} \right)$$
$$\forall \, v \in VM, s \in N \tag{34}$$

$$W_s^{(MF)} = \sum_{v \in VM} W_{v,s}^{(MFR)}$$
$$\forall \, s \in N \tag{35}$$

$$W_{v,p,s}^{(AFR)} = \delta_{v,p,s}^{(AF)} \, W_v$$
$$\forall \, v \in VM, s \in N, p \in P \tag{36}$$

$$W_{v,p,s}^{(AFR)} = \left( \frac{D_{v,p,s}^{(AF)}}{r_v \, x} \, M \, \delta_{v,p,s}^{(AF)} \right) + \left( W_v^{(R)} D_{v,p,s}^{(AF)} \right)$$
$$\forall \, v \in VM, s \in N, p \in P \tag{37}$$

$$W_{p,s}^{(AF)} = \sum_{v \in VM} W_{v,p,s}^{(AFR)}$$
$$\forall \, s \in N \tag{38}$$

Constraints (30), (33) and (36) calculate the workload of a VM replica in a cloud, a metro fog and an access fog, respectively under a constant workload profile. Constraints

(31), (34) and (37) calculate the workload of a VM replica in a cloud, a metro fog and an access fog, respectively as a linear function of the traffic resulting from serving users of the replica plus the workload baseline. Constraints (32), (35) and (38) calculate the total workload of a cloud, a metro fog and an access fog, respectively by summing the workload of VMs hosted in it.

Number of servers in cloud and fog constraints:

$$S_s^{(C)} \geq \frac{W_s^{(C)}}{S^{(maxW)}}$$
$$\forall\, s \in N \quad (39)$$

$$S_s^{(MF)} \geq \frac{W_{v,s}^{(MFR)}}{S^{(maxW)}}$$
$$\forall\, s \in N \quad (40)$$

$$S_{p,s}^{(AF)} \geq \frac{W_{p,s}^{(AF)}}{S^{(maxW)}}$$
$$\forall\, s \in N, p \in P \quad (41)$$

Constraints (39) - (41) calculate the number of servers in each cloud, metro fog and access fog, respectively based on the CPU utilization as the CPU draws the largest proportion of the server power consumption [55].

Number of router ports and switches in cloud and fog:

$$R_s^{(C)} \geq \frac{\sum_{v \in VM} \sum_{d \in N} D_{v,s,d}^{(C)}}{R^{(CB)}}$$
$$\forall\, s \in N \quad (42)$$

$$SW_s^{(C)} \geq \frac{\sum_{v \in VM} \sum_{d \in N} D_{v,s,d}^{(C)}}{SW^{(CB)}}$$
$$\forall\, s \in N \quad (43)$$

$$R_s^{(MF)} \geq \frac{\sum_{v \in VM} D_{v,s}^{(MF)}}{R^{(MFB)}}$$
$$\forall\, s \in N \quad (44)$$

$$SW_s^{(MF)} \geq \frac{\sum_{v \in VM} D_{v,s}^{(MF)}}{SW^{(MFB)}}$$
$$\forall\, s \in N \quad (45)$$

$$R_{p,s}^{(AF)} \geq \frac{\sum_{v \in VM} D_{v,p,s}^{(AF)}}{R^{(AFB)}}$$
$$\forall\, s \in N, p \in P \quad (46)$$

$$SW_{p,s}^{(AF)} \geq \frac{\sum_{v \in VM} D_{v,p,s}^{(AF)}}{SW^{(AFB)}}$$
$$\forall\, s \in N, p \in P \quad (47)$$

Constraints (42) - (47) calculate the number of routers ports and switches in each cloud, metro fog and access fog, respectively.

Number of metro router ports and ethernet switches in metro network constraints:

$$R_s^{(M)} \geq \frac{\sum_{v \in VM} \sum_{s \in N} D_{v,s,d}^{(C)} + \sum_{v \in VM} D_{v,s}^{(MF)}}{R^{(MB)}}$$
$$\forall\, s \in N \quad (48)$$

$$SW_s^{(M)} \geq \frac{\sum_{v \in VM} \sum_{s \in N} D_{v,s,d}^{(C)} + \sum_{v \in VM} D_{v,s}^{(MF)}}{SW^{(MB)}}$$
$$\forall\, s \in N \quad (49)$$

Constraints (48) and (49) calculate the number of routers ports and switches, respectively, in each metro network.

Traffic demand on IP over WDM core network constraint:

$$L_{s,d} = \sum_{v \in VM} D_{v,s,d}^{(C)}$$
$$\forall\, s, d \in N \quad (50)$$

Constraint (50) calculates the demand between the IP over WDM nodes by summing the demand due to VMs placed in the clouds.

Flow conservation constraint in the IP layer:

$$\sum_{j \in N: i \neq j} L_{i,j}^{s,d} - \sum_{j \in N: i \neq j} L_{i,j}^{s,d} = \begin{cases} L_{s,d} & i = s \\ -L_{s,d} & i = d \\ 0 & otherwise \end{cases}$$
$$\forall\, s, d, i \in N : s \neq d \quad (51)$$

Constraint (51) represents the flow conservation for IP layer on the IP over WDM network. It ensures that the total incoming traffic equal the total outgoing traffic in all nodes; excluding the source and destination nodes.

Virtual link capacity constraint:

$$\sum_{s \in N} \sum_{d \in N: s \neq d} L_{i,j}^{s,d} \leq C_{i,j}\, \mathcal{W}^{(B)}$$
$$\forall\, i, j \in N : s \neq d \quad (52)$$

Constraint (52) ensures that the traffic transmitted through a virtual link does not exceed its maximum capacity.

Flow conservation constraint in the optical layer:

$$\sum_{n \in Nm_m} \dot{w}_{m,n}^{i,j} - \sum_{n \in Nm_m} \mathcal{W}_{m,n}^{i,j} = \begin{cases} C_{i,j} & m = i \\ -C_{i,j} & m = j \\ 0 & otherwise \end{cases}$$
$$\forall\, i, j, m \in N : i \neq j \quad (53)$$

Constraint (53) represents the flow conservation for the optical layer. It ensures that the total number of incoming wavelengths in a virtual link is equal to the total number of outgoing wavelengths in all nodes excluding the source and destination nodes of the virtual link.

Physical link capacity:

$$\sum_{i \in N} \sum_{j \in N: i \neq j} \dot{w}_{m,n}^{i,j} \leq \dot{w}\, F_{m,n}$$
$$\forall\, m, n \in N \quad (54)$$

$$\dot{w}_{mn} = \sum_{i \in N} \sum_{j \in N: i \neq j} \dot{w}_{m,n}^{i,j}$$
$$\forall\, m, n \in N \quad (55)$$

Constraints (54) and (55) represent the physical link capacity limit. Constraint (54) ensures that the number of wavelengths in virtual links traversing a physical link does not exceed the maximum capacity of fibers in the physical link. Constraint (55) calculates the number of wavelengths in a physical link as the sum of wavelength channels in virtual links traversing the physical link.

Total number of router ports in a core node:

$$R_s^{(AC)} = \frac{1}{\mathcal{W}^{(B)}} \sum_{d \in N} L_{s,d}$$
$$\forall\, s \in N \tag{56}$$

$$R_s^{(AE)} = R^{(MR)} \left( \frac{1}{\mathcal{W}^{(B)}} \sum_{s \in N} L_{s,d} \right)$$
$$\forall\, d \in N \tag{57}$$

Constraint (56) calculates the total number of router ports in each core node that aggregate the traffic from/to the clouds. Constraint (57) calculates the total number of router ports in each core node that aggregate the traffic from/to edge routers.

### III. CLOUD-FOG ARCHITECTURE MILP MODEL RESULTS

In this section, the optimal VMs placement over AT&T distributed cloud architecture is investigated. The AT&T core networks topology is illustrated in Fig. 3 [56]. The AT&T core network consists of 25 nodes and 54 bidirectional links [56]. We consider an architecture where each core node is connected to two PON networks through a metro network consisting of a single ethernet switch and two metro routers (illustrated in Fig 3.1). The PON access network is considered to connect 512 different locations. The total capacity of each OLT is 1280 Gbps [57].

We start by considering the optimization of the placement a single VM as the simplest representative problem. Then we consider optimization in a realistic scenario with multiple VMs.

#### A. Simple Representative Scenario:

We investigate how the energy efficient placement of a single VM over cloud-fog architecture varies based on three factors; the CPU requirements, download traffic and PUE values. The impact of the VM workload profile on the VM placement is examined by considering constant and linear workload profiles. For the linear workload profile, a simple linear profile with no baseline is considered. The workload of a VM of a constant workload profile and the workload of a VM of a linear workload profile that serves the maximum number of users are considered. Three workloads: 10%, 50% and 100% of the server CPU capacity are considered. The users are considered to access VMs with one of following download rates; 0.1 Mbps, 1 Mbps, 10 Mbps, 20 Mbps, 50 Mbps, 100 Mbps or 200 Mbps. Each VM is considered to have 800 users. The PUE is a metric used to determine the total energy consumption required by the facility that hosts the clouds, fog nodes or network nodes. The total facility power consumption includes the power consumption of the computing and communication hardware, in addition to the power consumption due to IT cooling, lighting, etc. PUE is the ratio of this total power consumption to the IT (computing and communication) infrastructure power consumption. Based on the US data center energy usage [58], the PUE varies based on the datacentre size as more efficient cooling technologies are used in larger datacenters. For best practice datacenters, PUE of clouds, metro fogs and access fogs take values of 1.3, 1.4 and 1.5, receptively [58]. For datacenters from 2014, the PUE values considered are 1.7, 1.9 and 2.5, respectively [58]. In network infrastructures, a typical telecom office PUE value is 1.5 [35].

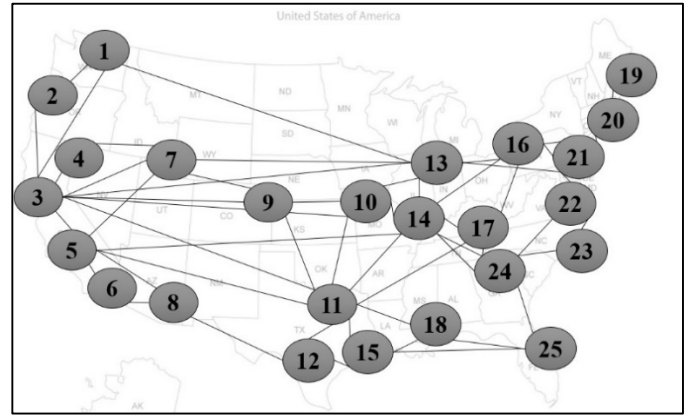

Figure 3: AT&T core network topology.

The Cisco Carrier Routing System 1 (CRS-1) [59] is considered as a core IP router. CRS-1 provides 160 Gbps routing capacity in 4 ports while consuming 2551W. Therefore, the power consumption of each 40 Gbps router port is 638W. Also, the Cisco NCS 5502 router [60] is considered as the cloud and metro networks router which consumes 30W per 40 Gbps port. In the metro and fog datacentre, Cisco NCS 5501 [60] is considered with a power consumption of 13W per 40 Gbps port. Furthermore, the Cisco Nexus 93180YC-EX [61] switch is considered as metro, cloud and metro fog LAN Ethernet switch with upload capacity of 600 Gbps and power rating at 470W. In access fog, the Cisco Nexus 93180YC-EX [61] switch is considered with capacity of 240 Gbps while consuming 210W. Tables I-III show the IP over WDM, metro and access network parameters and Table IV shows the Clouds and fogs parameters.

The MILP model is solved using the CPLEX solver over the University of Leeds high-performance computer (Polaris) using 16 nodes (256 cores) with 16 GByte of RAM per core. Each node comprises two eight-core Intel 2.6 GHz Sandy Bridge E5-2670 processors [62].

TABLE I
IP OVER WDM CORE NETWORK INPUT PARAMETERS OF THE MODEL

| | |
|---|---|
| Router port power consumption ($R^{(P)}$) | 638 Watt [59] |
| Transponder power consumption ($t^{(P)}$) | 129 Watt [63] |
| Regenerator power consumption ($G^{(P)}$) | 114 Watt, reach 2000 km [64] |
| EDFA power consumption ($e^{(P)}$) | 11 Watt [65] |
| Optical switch power consumption ($SW^{(P)}$) | 85 Watt [66] |
| Number of wavelengths in a fiber ($\mathcal{W}$) | 32 [67] |
| Bit rate of each wavelength ($\mathcal{W}^{(P)}$) | 40 Gbps [67] |
| Span distance between two EDFAs ($S$) | 80 km [65] |
| Network power usage effectiveness ($n$) | 1.5 [33] |

TABLE II
METRO NETWORK INPUT PARAMETERS OF THE MODEL

| | |
|---|---|
| Metro router redundancy ($R^{(MR)}$) | 2 |
| Metro edge router port bit rate ($R^{(MB)}$) | 40 Gbps |
| Metro edge router port power consumption ($R^{(MP)}$) | 30 Watt [60] |
| Metro ethernet switch bit rate ($SW^{(MB)}$) | 600 Gbps [61] |
| Metro ethernet switch power consumption ($W^{(MP)}$) | 470 Watt [61] |

TABLE III
ACCESS NETWORK INPUT PARAMETERS OF THE MODEL

| | |
|---|---|
| Number of PON networks in a node ($P$) | 2 |
| Number of users of VM services in each PON based on VMs popularity groups ($U_{v,p,d}$) | 13,000 users in each PON, six VMs popularity groups; 16%, 5%, 2%, 1%, 0.5% and 0.05%. |
| Maximum number of users of a single VM ($x$) | 800 concurrent users |
| Number of ONU devices in a PON network ($ONU_{p,d}^{(N)}$) | 512 |
| Power consumption of ONU device ($ONU^{(P)}$) | 5 Watt [68] |
| Number of OLTs in a PON network ($OLT_{p,d}^{(N)}$) | 1 |
| OLT Capacity ($OLT_{p,d}^{(B)}$) | 1280 Gbps [57] |
| OLT Power consumption ($OLT^{(P)}$) | 1842 W [57] |

TABLE IV
CLOUD AND FOG INPUT PARAMETERS OF THE MODEL

| | |
|---|---|
| Number of VMs ($V$) | 1 |
| User download rate ($r_v$) | {0.1, 1, 10, 20, 50, 100 or 200 Mbps} |
| Maximum workload of VM ($W_v$) | 10%, 50% and 100% |
| Server power consumption ($S^{(P)}$) | 333 Watt [69] |
| Maximum server workload ($S^{(maxW)}$) | 100% |
| Cloud and metro fog switch bit rate ($SW^{(CB)}, SW^{(MFB)}$) | 600 Gbps [61] |
| Cloud and metro fog switch power consumption ($SW^{(CP)}, SW^{(MFP)}$) | 470 Watt [61] |
| Access fog switch bit rate ($SW^{(AFB)}$) | 240 Gbps [61] |
| Access fog switch power consumption ($SW^{(MFP)}$). | 210 Watt [61] |
| Cloud and fog switch redundancy ($SW^{(R)}$) | 2 |
| Cloud and fog router port bit rate ($R^{(CB)}, R^{(MFB)}, R^{(AFB)}$) | 40 Gbps |
| Cloud router port power consumption ($R^{(CP)}$) | 30 Watt [60] |
| Metro and access fog router port power consumption ($R^{(MFP)}, R^{(AFP)}$) | 13 Watt [60] |
| Cloud power usage effectiveness ($c$) | 1.3 or 1.7 [58] |
| Metro fog power usage effectiveness ($m$) | 1.4 or 1.9 [58] |
| Access fog power usage effectiveness ($a$) | 1.5 or 2.5 [58] |

Figs. 4 (a), (b) and (c) show the optimal placement of VMs of 10%, 50% and 100% CPU requirements, respectively, considering the best practice PUE values. In each figure, the x-axis is the VM workload profile, the y-axis is the data rates which range from 0.1 Mbps to 200 Mbps and the z-axis is the percentage of VM replicas in each location over the cloud-fog architecture.

The placement of VM with linear workload profile is not affected by the VM workload as serving users will consume the same power whether centralized in a single VM or distributed among multiple replicas with smaller workloads. However, the higher PUE of fog nodes compared to the cloud, results in a situation where distributing replicas into fog processing nodes incurs additional power consumption as the PUE value of fog nodes is higher than that of clouds. Hence, there is a trade-off between the network power saved by replicating VMs into fog nodes and the additional power consumed by these replicas. The creation of a VM replica results in power savings if the former power exceeds the latter power. At data rates of 1 Mbps and higher, VMs of 10%, 50% and 100% workloads are offloaded to access fog processing nodes considering a linear workload profile.

For constant workload profile, replicas are less energy efficient, therefore, offloading VMs to fog nodes decreases as the VM workload increases. While VMs of 10% workload and 20 Mbps are fully offloaded to metro fogs, 50% and 100% workload VMs are replicated only to clouds. Also, users of VMs of 50% workload at 100 Mbps data rate as well as VMs of 100% workload at 200 Mbps data rate are served by clouds and metro fog nodes. A VM replica is offloaded to 14 metro fog nodes (in core nodes 1, 2, 4, 6, 7, 8, 13, 16, 19, 20, 21, 22, 23, 25) while users from other nodes are served by the replica placed in the cloud in core node 11 which they can access by traversing a single hop in the core network. These 14 metro fog nodes are selected to host replicas of the VM as the traffic flows traverse more than a single hop in the IP over WDM network to access the VM placed in the cloud hosted in node 11 and therefore increase the need for IP router ports (the most power consuming device in the IP over WDM network).

The results also show that VMs with higher data rates justify the creation of more replicas closer to user premises in the fog layer. Thus, the power consumption of the network, which is the major contributor to the power consumption in the cloud-fog architecture, is reduced. For example, VMs of 10% workload under the linear workload profile, are fully replicated to clouds and offloaded to access fog nodes for VMs of 0.1 Mbps, and ≥ 1 Mbps user data rates, respectively.

Placing VMs in cloud architecture with higher PUE (2014 PUE), as in Fig. 5, increases the replicas power consumption and therefore limits offloading VMs into the fog processing nodes, e.g., VM of constant workload profile with 100% workload and 200 Mbps data rate, that are fully offloaded to metro fogs considering clouds of best practice PUE, are limited to clouds with 2014 PUE.

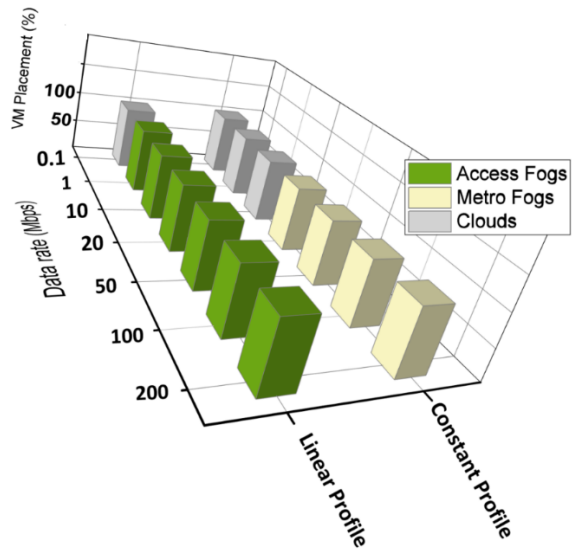

(a)

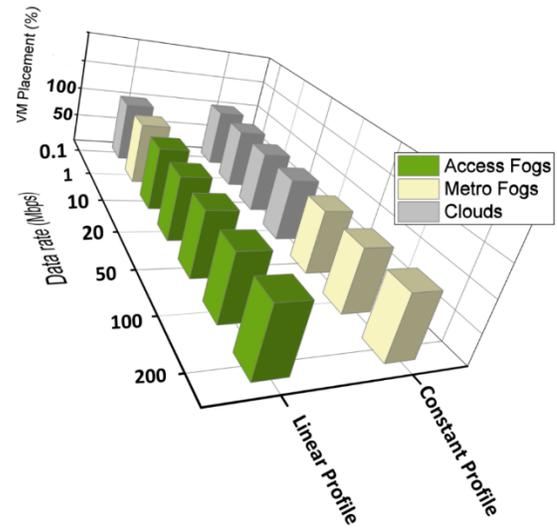

(a)

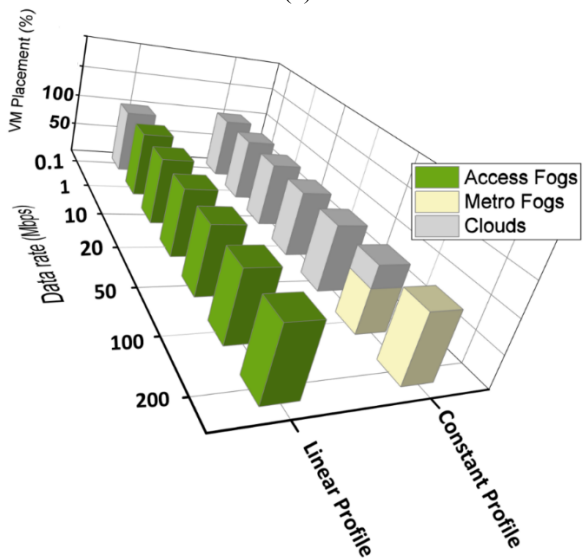

(b)

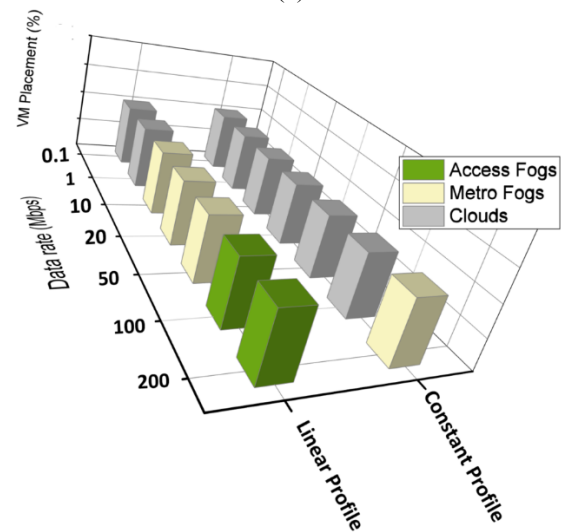

(b)

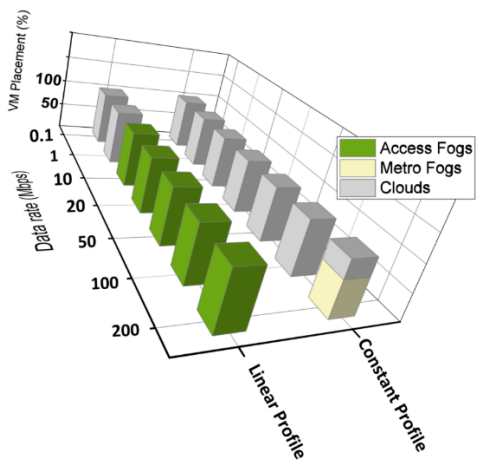

(c)

Figure 4: Optimal VM placement of (a) constant profile at 10% of CPU and linear profile with peak utilization at 10%, (b) 50% case, (c) 100% case at different data rates considering best practice PUE value ($c=1.3, m = 1.4, a = 1.5$).

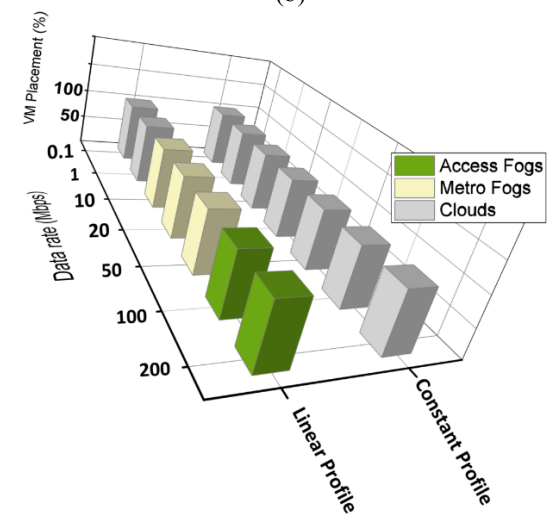

(c)

Figure 5: Optimal VM placement of (a) constant profile at 10% of CPU and linear profile with peak utilization at 10%, (b) 50% case, (c) 100% case at different data rates considering 2014 PUE value ($c=1.7, m = 1.9, a = 2.5$).

## B. Realistic Scenario

In this scenario, analysis based on realistic number of users and VM popularity is studied. According to Cisco Visual Network Index (VNI) [70], in 2016, the average broadband data rate in US was 36 Mbps. Therefore, each OLT is assumed to be able to serve ~35k connections (or users). Cisco VNI also reports that 76% of all Internet traffic crossed clouds in 2016. SimilarWeb [71], an online tool which provides Internet traffic statistics and analytics, shows that the top 300 applications or websites have a 50% share of all traffic. Accordingly, 13k users are considered in each PON (~50% of clouds traffic, i.e. 38% of the total traffic) to access the VMs hosting the top 300 applications or websites. The popularity of these VMs is considered to follow a Zipf distribution [72]. To simplify the analysis, VMs' popularity is divided into 6 groups as follows; 16%, 5%, 2%, 1%, 0.5% and 0.05% of the total users. The number of VMs in each popularity group are 1, 3, 5, 16, 65 and 210, respectively.

Each VM is assumed to require 50% of the CPU's server capacity in order to serve 800 users. Based on the literature [44] - [46], [73], [74], in such a case, a VM can serve 800 users with low error rate. VMs of a linear workload are considered to have a workload baseline of 1%, 5% or 40% of the total server CPU capacity based on the CPU requirements for state of the art applications [44], [46], [73] (e.g. 1% workload baseline for database applications, 5% for website applications, and 40% for video games and web conference applications). The users are considered to access the VMs with one of the following data rates; 1 Mbps (low), 10 Mbps (medium) or 25 Mbps (high). Such data rates represent the recommended download speed to access the content of the state of the art applications, e.g. 1 Mbps for light web browsing [75] (emails, Google docs [76] and websites with lower definition video content [77]), 10 Mbps for applications processing high-definition video quality [78] and online multiplayer games [79], and 25 Mbps for applications processing ultra-high video quality [80].

The optimized VMs placement over the cloud-fog architecture, referred to as *Optimized clouds and fogs placements (OC&F)* approach, is compared to the *Optimized clouds* (OC) approach where VMs are optimally placed in clouds distributed over the core network and *AT&T clouds* (ATT) where the VMs are placed in nodes 1, 3, 5, 6, 8, 11, 13, 17, 19, 20, 22, and 25 according to AT&T datacenters map [56].

In addition to the parameters in Table I to Table IV, Table V shows the additional/modified parameters considered for the following results.

TABLE V
I INPUT PARAMETERS USED IN THE MODEL

| | |
|---|---|
| Router port power consumption ($R^{(P)}$) | 638 Watt [59] |
| Transponder power consumption ($t^{(P)}$) | 129 Watt [63] |
| Regenerator power consumption ($G^{(P)}$) | 114 Watt, reach 2000 km [64] |
| EDFA power consumption ($e^{(P)}$) | 11 Watt [65] |
| Optical switch power consumption ($SW^{(P)}$) | 85 Watt [66] |
| Number of wavelengths in a fiber ($\dot{w}$) | 32 [67] |
| Bit rate of each wavelength ($W^{(P)}$) | 40 Gbps [67] |
| Span distance between two EDFAs ($S$) | 80 km [65] |

### 1) *Linear Workload Profile (1% Workload Baseline):*

Fig. 6 shows the power consumption resulting from placing VMs of 1% minimum CPU workload considering the different placement approaches under 1, 10 and 25 Mbps user data rates. The efficiency of VMs with 1% minimum CPU workload allows the creation of more efficient VMs replicas as the workload is proportional to the number of users served by the VM with a trivial minimum workload required by each VM. Under 1 Mbps data rate, the OC&F approach achieves 6% reduction in the total power consumption compared to the AT&T clouds. The total reductions mount to 40% under 10 Mbps data rate and 64% under 25 Mbps data rate. The savings achieved by the OC&F approach compared to the OC approach are 4%, 31% and 48% under the low, medium and high data rates, respectively. Also, the savings achieved by the OC approach compared to the ATT approach are 2%, 9% and 16% under the low, medium and high data rates, respectively.

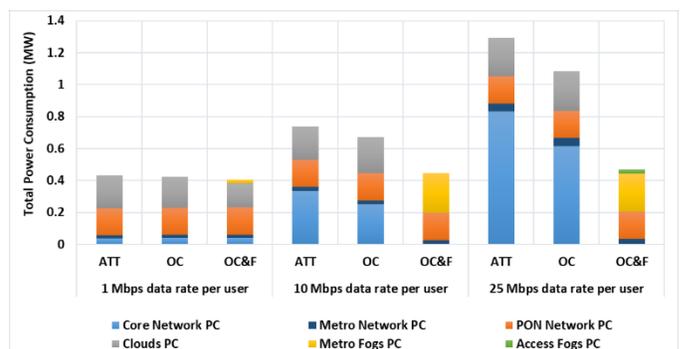

Figure 6: the power consumption of different VMs placement approaches considering VMs of 1% minimum CPU workload.

In Fig. 7 and Fig. 8, we further investigate the OC and OC&F placement approaches by looking at how VMs of different data rates and popularity groups are placed considering the low, medium and high data rates. Note that the different colors show if a VM of a certain popularity is placed in this location or not, i.e. it does not represent the number of replicas.

Fig. 7 shows the optimal VMs placement under the OC approach. Note that the different colors indicate the creation of a replica of the VM in the cloud, i.e. not the number of replicas. The efficiency of VMs has allowed the creation of multiple replicas as the workload is proportional to the number of users served by the VM with a limited workload baseline. The efficient workload profile of the VMs has justified the replication of VMs of popularity greater than 0.5% into 10 clouds for 1 Mbps data rates and into 25 clouds (full replication) for 10 Mbps data rates. VMs of 0.05% popularity are only replicated into 2 clouds. The high traffic of VMs of 25 Mbps data rate allows full replication for the different popularity groups across all clouds.

Figure 8(a) shows that VMs with a low user data rate of 1 Mbps have only justified creating three metro fogs in nodes 6, 8, and 19 as the traffic flows from these nodes traverse more than a single hop in the IP over WDM network to access the replicas optimally placed in the distributed clouds built in nodes 3, 11, 20, and 24. Thus, these fog nodes are built to serve the user demands locally, and consequently, eliminate the need for IP router ports. However, VMs with the lowest popularity (0.05%) have only justified the creation of two replicas in nodes 11 and 20.

VMs with 10 Mbps data rate are fully offloaded to every metro fog as shown in Fig. 8(b). Note that, VM users are uniformly distributed across the metro and access networks, thus, the placement of a VM is consistent across all the metro fog nodes. In Fig. 8(c), VMs with a high data rate of 25 Mbps show full replication in metro fog nodes, VMs with 16% popularity justified creating VM replicas in some access fog nodes. Although, we are able to reduce the traffic traversing the metro network and consequently reduce the total power consumption, however, VMs with 16% popularity are not fully replicated to access fog nodes. There are a number of replicas offloaded to metro fog nodes. The reason for that is the on-off power consumption profile of fog and network resources. Thus, before creating a new fog node in the access network, VMs are consolidated into the available resources that remain from the placement of other VMs that share the same architecture.

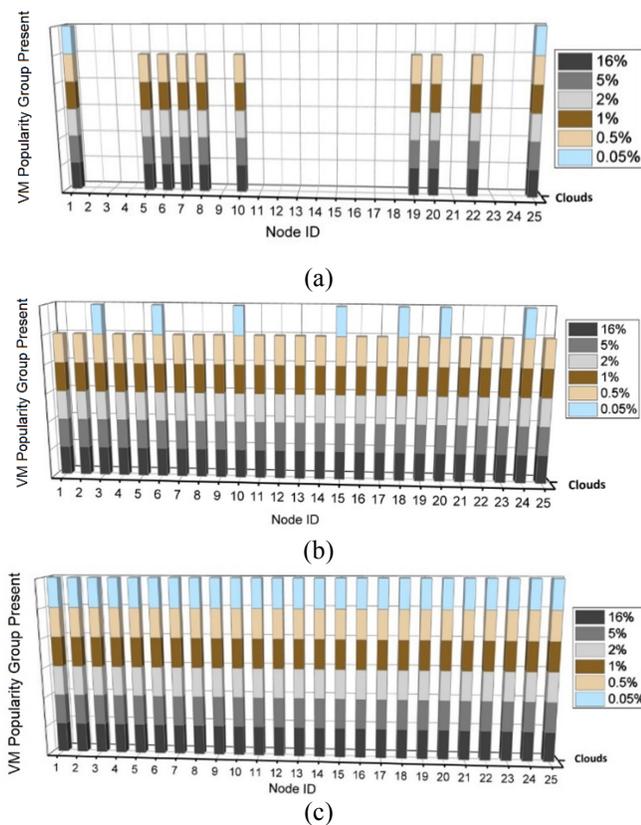

Figure 7: Optimal placement of different VMs popularity groups of 1% workload baseline under the OC approach with (a) 1 Mbps data rate per user, (b) 10 Mbps data rate per user and (c) 25 Mbps data rate per user.

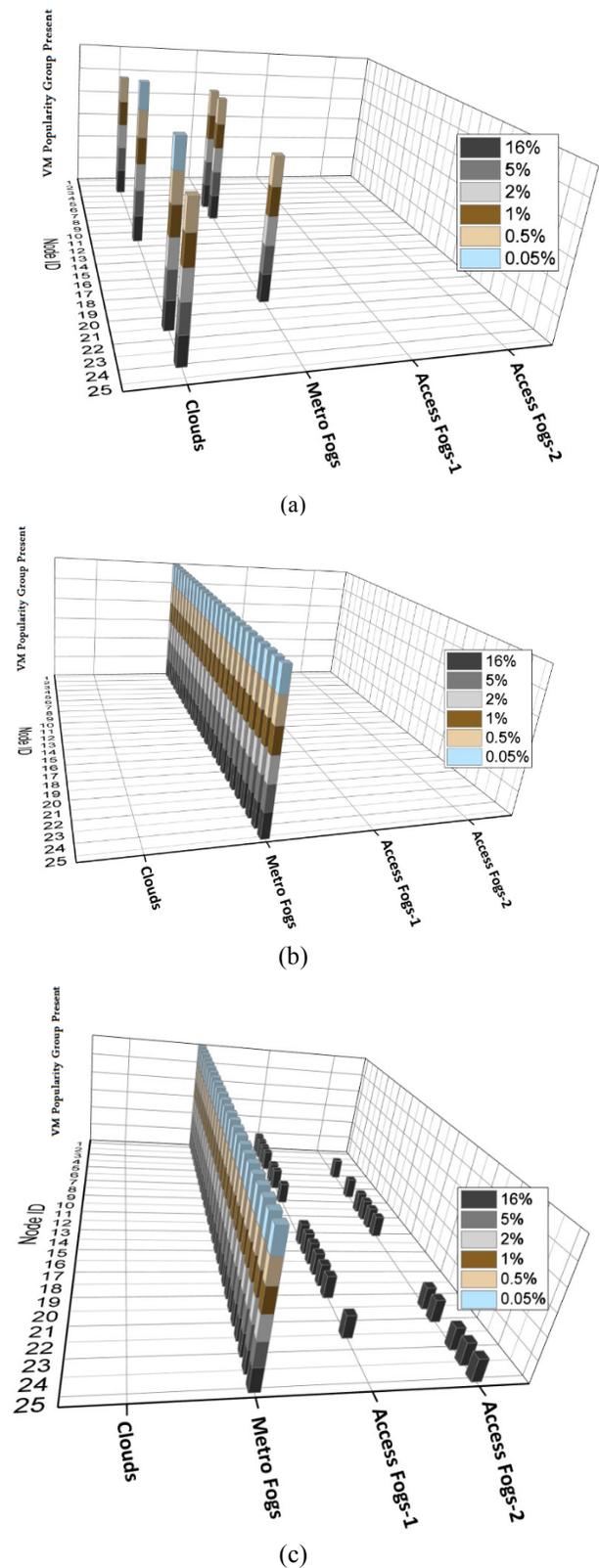

Figure 8: Optimal placement of different VMs popularity groups of 1% workload baseline under the OC&F approach with (a) 1 Mbps data rate per user, (b) 10 Mbps data rate per user and (c) 25 Mbps data rate per user.

In Fig. 9, OC&F1 and OC&F2 placement approaches are introduced. The former represents the optimal placement considering clouds and metro fog nodes only and the latter shows the optimal placement considering the three computing layers; clouds, metro and access fogs. These two approaches show how introducing fog nodes in the access network (OC&F2), in addition to metro fogs, is able to save in terms of total power consumption compared to an approach that considers only fog nodes connected to metro network (OC&F1). Under a 25 Mbps user data rate, it can be observed that the OC&F2 approach achieves 6% extra power saving compared to the OC&F1 approach.

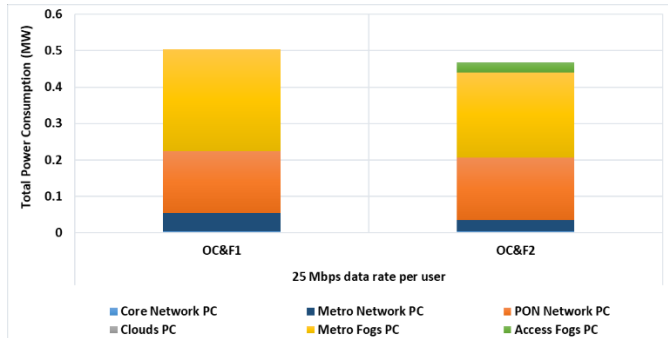

Figure 9: the power consumption considering OC&F1 and OC&F2 placement approaches. OC&F1 represents the optimal placement considering clouds and metro fogs only and OC&F2 represents the optimal placement considering clouds, metro and access fogs.

Fig. 10 and Fig. 11 show the number of servers required to host VM replicas under the OC and OC&F approaches, respectively. The number of servers is a function of the number of VM replicas hosted and their workload. For instance, the OC&F approach under 25 Mbps user data rate (Fig. 9(c)) requires 18 servers in each metro fog and 2 servers in access fogs to host VM replicas. Such a number of servers can be practically attached to the metro edge routers to create the metro fog layer and to the OLT in the access network to create the access fog layer.

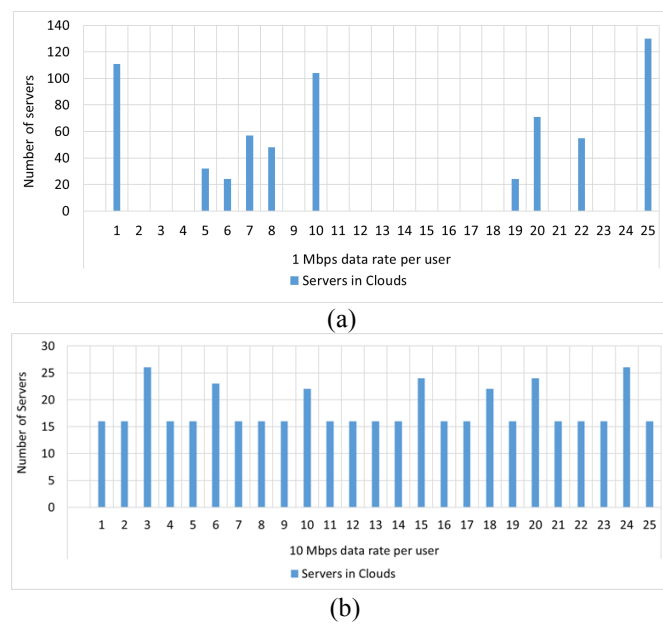

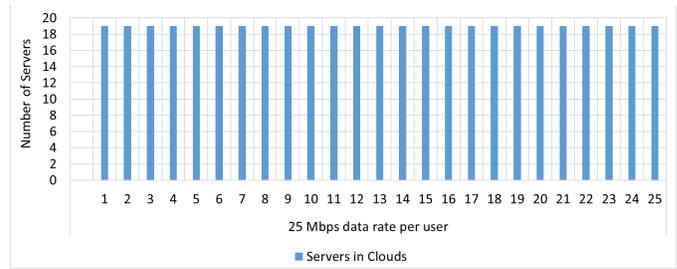

(c)

Figure 10: Number of servers in OC approach required to host VMs of 1% workload baseline with (a) 1 Mbps data rate per user (b) 10 Mbps data rate per user (c) 25 Mbps data rate per user.

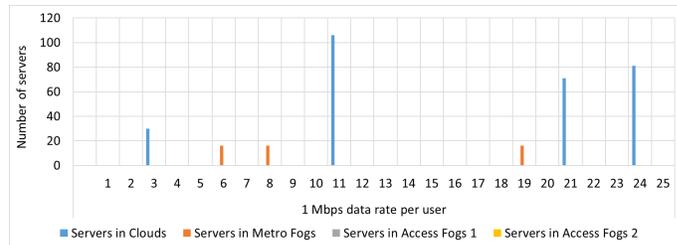

(a)

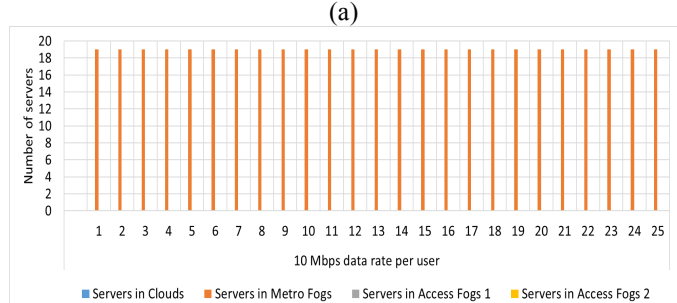

(b)

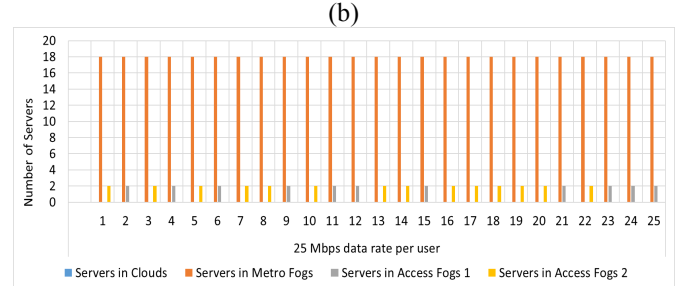

(c)

Figure 11: Number of servers in OC&F approach required to host VMs of 1% workload baseline with (a) 1 Mbps data rate per user (b) 10 Mbps data rate per user (c) 25 Mbps data rate per user.

2)    *5% Minimum CPU Workload:*

Fig. 12 shows the power savings achieved under VMs that have a linear workload profile with a 5% minimum CPU utilization. Increasing the minimum CPU utilization of the VM workload profile to the current value of 5% reduces the efficiency of creating more VMs replicas. The total savings achieved under the OC&F approach compared to the AT&T cloud are 12%, 35% and 55% under the low, medium and high data rates, respectively. Compared to the OC approach, there is no extra power saving achieved under low data rate, as the total traffic has not justified replicating any VM into fogs. Under the medium and high user data rates, the power savings achieved are 28% and 47%, respectively.

Figs. 12 (a) and (b) illustrate the placements of the VMs of 5% minimum CPU utilization considering the OC&F placement approach under low and high user data rates, respectively. VMs with low user data rates are dispersed among distributed clouds. The low user data rates have not justified offloading VMs to any fog node. VMs of ≥ 1% popularity have justified the creation of five cloud locations. VMs with 0.5% and 0.05% popularity have only justified the creation of three and two replicas, respectively. Under the high user data rates, it can be observed that VMs with ≥ 0.5% and ≤ 5% popularity are fully offloaded to the metro fogs. In addition, VMs with 16% popularity have justified the creation of replicas in some access fogs. Whereas, VMs with 0.05% popularity have only justified the creation of two replicas in nodes 3 and 14.

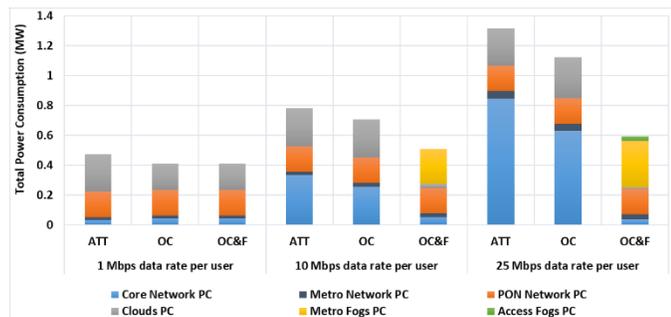

Figure 12: The power consumption of different VMs placement approaches considering VMs of 5% workload baseline.

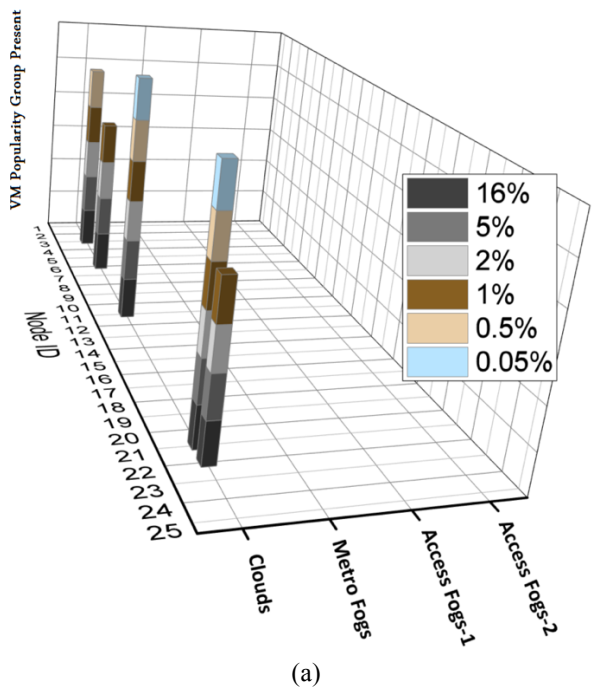

(a)

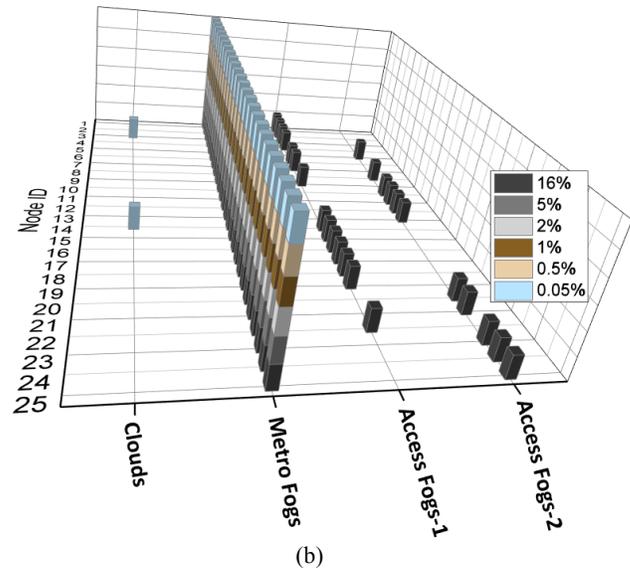

(b)

Figure 13: the optimal placement of different VMs popularity groups of 5% minimum CPU workload under the OC&F approach with 25 Mbps data rate per user.

### 3) *40% Minimum CPU Workload:*

Fig. 14 shows the power savings achieved under VMs of 40% minimum CPU utilization. The total savings achieved under the OC&F approach compared to the AT&T clouds are 53%, 44% and 48% under the low, medium and high user data rates, respectively. Compared to the OC, there is no extra power saving achieved under the low user data rates, as the total traffic has not justified replication of any VM into any fog node. Under the medium and high user data rates, the power savings achieved are 12% and 31%, respectively.

Figs. 15 (a), (b) and (c) illustrate the optimal VMs placement under low, medium and high user data rates, respectively, considering the OC&F approach. It can be observed that increasing the minimum CPU utilization of VM workload to 40% reduces the efficiency of creating more replicas of VMs with a low popularity across distributed cloud and fog nodes, compared to VMs with 1% or 5% minimum CPU utilization. VMs with data rate of 1 Mbps are replicated among distributed clouds. The low user data rates have not justified offloading VMs to any fog node. VMs of ≥ 1% popularity have justified the creation of five cloud locations. However, VMs with 0.5% and 0.05% popularity groups have only justified the creation of three and one replicas, respectively. Under medium user data rates, VMs with ≥ 1% popularity are offloaded to metro fogs whereas other popularity groups are optimally placed in clouds. Under high user data rates, despite the high workload baseline, VMs with high popularity of 16% justified the creation of VM replicas in some access fog nodes. VMs with ≥ 0.5% and ≤ 5% popularity are fully offloaded to metro fogs whereas VMs with 0.05% popularity have not justified the creation of multiple replicas. Only a single replica is optimally placed in node 11 to serve its distributed users.

Fig. 16 shows the number of servers required to host VM replicas under the OC&F approach with 25 Mbps data rate per user. The number of servers is a function of the number of VM replicas hosted and their workload.

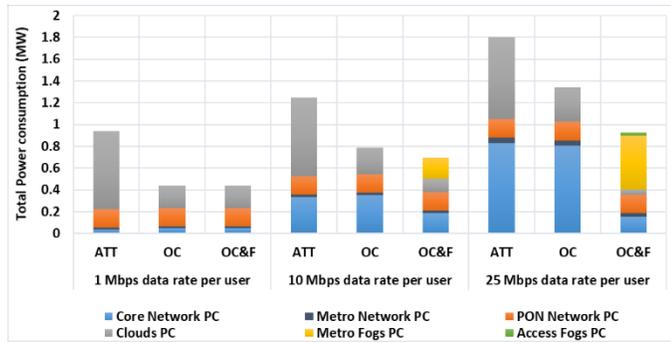

Figure 14: The power consumption of different VMs placement approaches considering VMs of 40% minimum CPU workload.

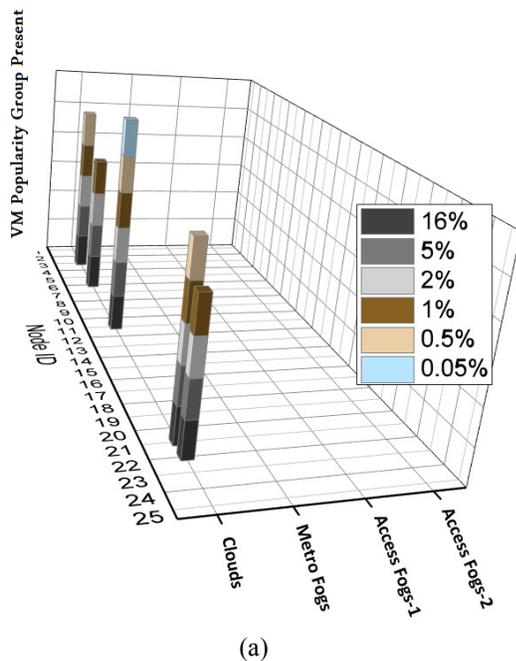

(a)

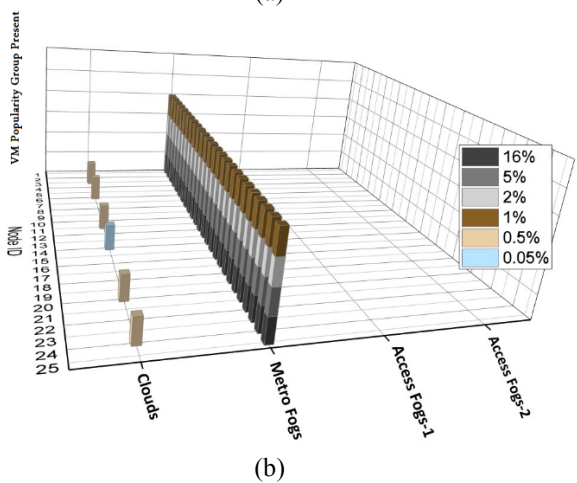

(b)

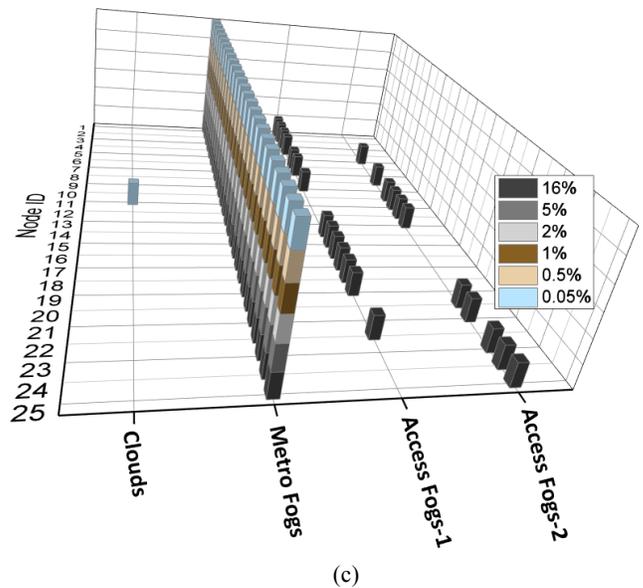

(c)

Figure 15: Optimal placement of different VMs popularity groups of 40% workload baseline under the OC&F approach with (a) 1 Mbps data rate per user, (b) 10 Mbps data rate per user and (c) 25 Mbps data rate per user.

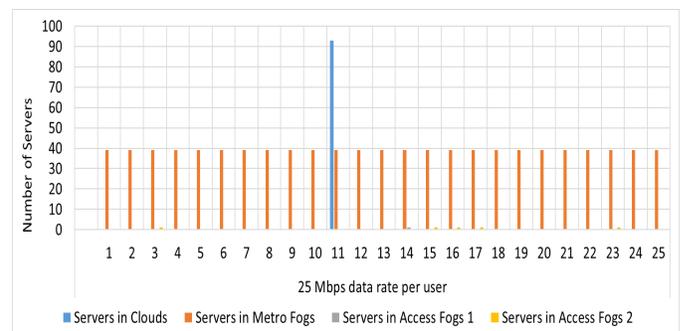

Figure 16: Number of servers required to host VMs of 40% minimum CPU workload under the OC&F approach with 25 Mbps data rate per user.

## IV. ENERGY EFFICIENT VM PLACEMENT HEURISTIC FOR THE CLOUD AND FOG

The VM placement problem over cloud-fog architecture is a nondeterministic polynomial (NP)-hard problem. For example, if $v$ is the number of VMs and $s$ is the number of servers, then the number of possible VM placements in different servers is $vs$. In case of replicating VMs into multiple data centres ($N$), exhaustive search of data centre distributed locations require the evaluation of $\left(\sum_{i=1}^{N} \frac{N!}{(N-i)!}\right)$ placement combinations in order to find the optimal number and locations of VM copies needed. Therefore, it is not practical to apply the MILP model in a real time large implementation. Heuristics can provide simple and fast operation in real time that may approach that of the optimal MILP solution. The optimal solutions obtained from the MILP model can thus offer a benchmark for determining the performance of the heuristics developed. A supervised learning algorithm is adopted here to develop a heuristic solution. Supervised learning is a branch of machine learning where an input is matched with an output based on a sample of input-output pairs. VMs are classified into different types based on user download rates, VM workloads and VM popularity. The optimum placement of different VM types are found in an offline phase. VMs are matched to their type in real time (online phase) and placed according to the placement obtained in the offline phase. In this section, we develop a real-time implementation of the MILP model, referred to as *energy efficient VM placement heuristic for the cloud-fog architecture* (EEVM-CF), to mimic the MILP model. The EEVM-CF heuristic consists of two-phases: offline phase and online phase.

In the offline phase, as shown in the flowchart in Fig. 17(a), VMs are classified into different types and the optimal placement of different VMs types are found. The search space, P, to find the optimum placement for each VM type includes the most energy efficient placement to place 1 replica, 2 replicas… up to $N$ replicas in the clouds, where $N$ is the number of clouds. For fog nodes, there are two placement scenarios. In one scenario, VMs are replicated to the metro fog and in all core nodes and in the other scenario VMs are replicated to the two access fog nodes in all core nodes. The traffic resulting from replicating the VMs in clouds and fogs and the workload of VMs of a linear workload profile are calculated based on the number of users each VM serves. The networks, the cloud and the fog power consumptions are calculated and the optimum placement of a VM type is the placement that results in the minimum total power consumption.

Then, VMs are matched to their type in real time (online phase), which is shown in the flowchart in Fig. 17(b) and placed according to the placement obtained in the offline phase. Then, the traffic resulting from replicating the VM in the cloud is routed on core network based on minimum hop routing [53] and the workload of clouds where the VM replicas are placed is updated. After placing all VMs, the total power consumption of the distributed cloud is calculated. After placing all VMs, the total power consumption of cloud-fog architecture is calculated.

The heuristics are examined by considering the AT&T network as a network example. The EEVM-CF heuristic took 55 seconds to evaluate the offline phase and 2 seconds to evaluate the online phase running on Intel i-5 core machine with 16GB RAM. Fig. 18 compares the total power consumption of EEVM-CF with the MILP model considering the network, cloud and fog parameters discussed in Section III. The heuristic is evaluated under 1%, 5% and 40% workload baseline considering 1 Mbps, 10 Mbps and 25 Mbps user data rates. The results show that the gap between the EEVM-CF and the MILP model ranges between 1% and 2% of the total power consumption.

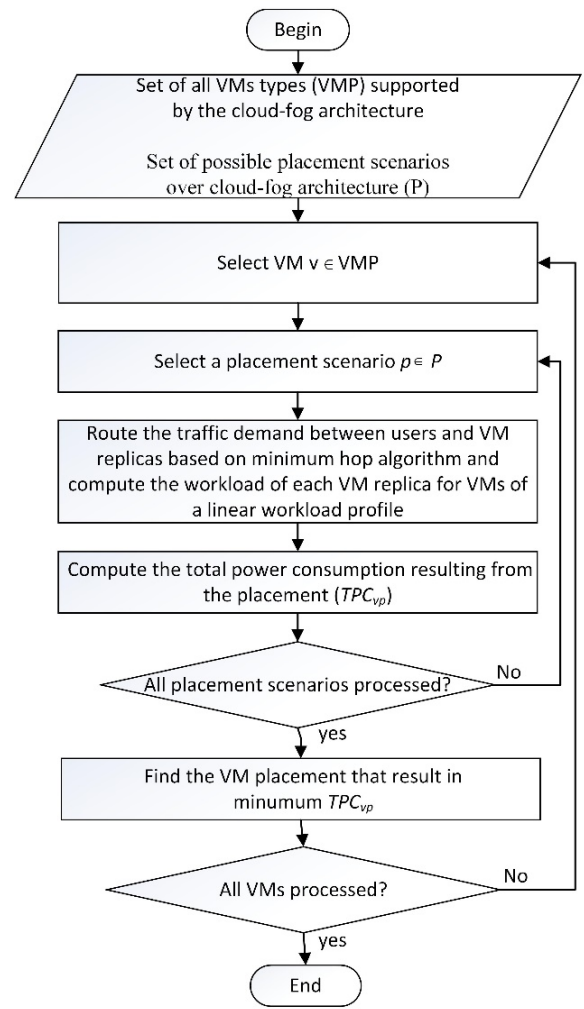

Figure 17: Flowchart of (a) the offline phase and (b) the online phase of EEVM-CF heuristic.

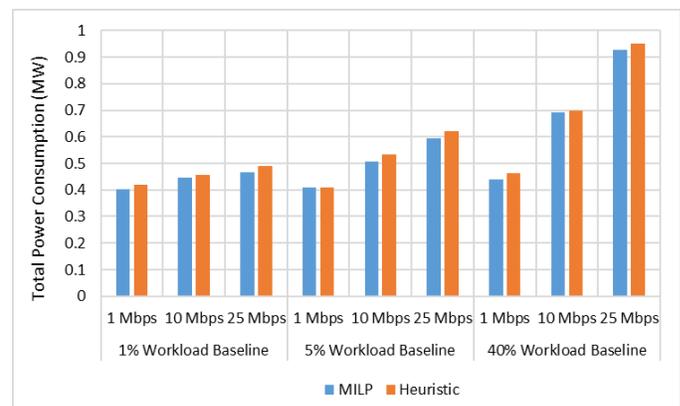

Figure 18: Total power consumption of the MILP model compared with EEVM-CF heuristics considering VMs with 1%, 5% and 40% CPU workload baseline.

## V. CONCLUSIONS

In this paper, the placement of VMs over a cloud-fog architecture is investigated with the aim of minimizing the total power consumption. The optimization is performed using a Mixed Integer Linear Programming (MILP) model considering AT&T and BT networks as use case scenarios. The MILP model is used to analyze the impact of different factors including VM popularity, the traffic between the VM and its users, the VM workload, the profile of the workload versus number of users, the proximity of fog nodes and the PUE.

The decision to serve users from fog nodes is driven by the trade-off between the network power saved by placing VMs in fog nodes close to end users, and the increase in processing power that results from replicating VMs to the fog. Our results demonstrate that VM placement in fog computing might lead to power saving depending on many factors which include workload and network bandwidth requirements of VMs, VMs popularity among users and the energy efficiency of distributed clouds.

The results evaluate a range of boundary and typical scenarios. For example, the processing power consumption of VMs of a linear workload profile with high data rate and minimum CPU utilization of 1% allows offloading VMs with 16% popularity to the access fog nodes. Other VMs are optimally replicated to metro fog nodes. Significant power savings of 48% compared to optimized placement in distributed clouds and 64% compared to a placement considering traditional cloud locations, have resulted from this offloading. VMs with linear workload and a minimum CPU utilization of 40% tend to offload fewer replicas into fog nodes as the high workload baseline means that VM consolidation in fewer locations is the most efficient approach.

Furthermore, we have developed a heuristic based on an offline exhaustive search, referred to as energy efficient VM placement heuristic for the cloud-fog architecture (EEVM-CF) to place VMs over the cloud-fog architectures in real-time. The heuristic results closely approach those of the MILP model.

## VI. ACKNOWLEDGEMENTS

The authors would like to acknowledge funding from the Engineering and Physical Sciences Research Council (EPSRC), INTERNET (EP/H040536/1), STAR (EP/K016873/1) and TOWS (EP/S016570/1) projects. All data are provided in full in the results section of this paper.

## Biographies

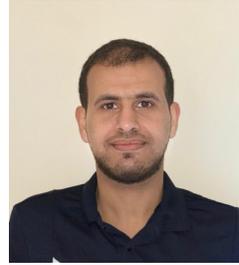

**Hatem A. Alharbi** received the B.Sc. degree in Computer Engineering (Hons.) from Umm Alqura University, Makkah, Saudi Arabia in 2012, the M.Sc. degree in Digital communication networks (with distinction) from University of Leeds, United Kingdom, in 2015. He is currently pursuing the Ph.D. degree with the School of Electronic and Electrical Engineering, University of Leeds, UK. He is currently a Lecturer in Computer Engineering department in the School of Computer Science and Engineering, University of Taibah, Saudi Arabia.

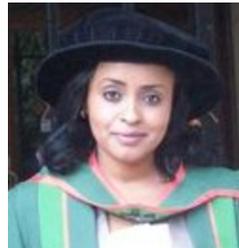

**Taisir EL-Gorashi** received the B.S. degree (first-class Hons.) in electrical and electronic engineering from the University of Khartoum, Khartoum, Sudan, in 2004, the M.Sc. degree (with distinction) in photonic and communication systems from the University of Wales, Swansea, UK, in 2005, and the PhD degree in optical networking from the University of Leeds, Leeds, UK, in 2010. She is currently a Lecturer in optical networks in the School of Electrical and Electronic Engineering, University of Leeds. Previously, she held a Postdoctoral Research post at the University of Leeds (2010– 2014), where she focused on the energy efficiency of optical networks investigating the use of renewable energy in core networks, green IP over WDM networks with datacenters, energy efficient physical topology design, energy efficiency of content distribution networks, distributed cloud computing, network virtualization and Big Data. In 2012, she was a BT Research Fellow, where she developed energy efficient hybrid wireless-optical broadband access networks and explored the dynamics of TV viewing behavior and program popularity. The energy efficiency techniques developed during her postdoctoral research contributed 3 out of the 8 carefully chosen core network energy efficiency improvement measures recommended by the GreenTouch consortium for every operator network worldwide. Her work led to several invited talks at GreenTouch, Bell Labs, Optical Network Design and Modelling conference, Optical Fiber Communications conference, International Conference on Computer Communications, EU Future Internet Assembly, IEEE Sustainable ICT Summit and IEEE 5G World Forum and collaboration with Nokia and Huawei.

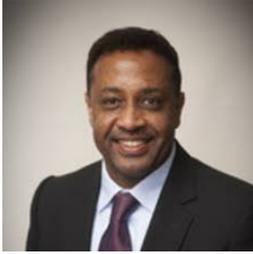

**Jaafar M. H. Elmirghani** is the Director of the Institute of Communication and Power Networks within the School of Electronic and Electrical Engineering, University of Leeds, UK. He joined Leeds in 2007 and prior to that (2000–2007) as chair in optical communications at the University of Wales Swansea he founded, developed and directed the Institute of Advanced Telecommunications and the Technium Digital (TD), a technology incubator/spin-off hub. He has provided outstanding leadership in a number of large research projects at the IAT and TD. He received the Ph.D. in the synchronization of optical systems and optical receiver design from the University of Huddersfield UK in 1994 and the DSc in Communication Systems and Networks from University of Leeds, UK, in 2014. He has co-authored Photonic switching Technology: Systems and Networks, (Wiley) and has published over 500 papers. He has research interests in optical systems and networks. Prof. Elmirghani is Fellow of the IET, Fellow of the Institute of Physics and Senior Member of IEEE. He was Chairman of IEEE Comsoc Transmission Access and Optical Systems technical committee and was Chairman of IEEE Comsoc Signal Processing and Communications Electronics technical committee, and an editor of IEEE Communications Magazine. He was founding Chair of the Advanced Signal Processing for Communication Symposium which started at IEEE GLOBECOM'99 and has continued since at every ICC and GLOBECOM. Prof. Elmirghani was also founding Chair of the first IEEE ICC/GLOBECOM optical symposium at GLOBECOM'00, the Future Photonic Network Technologies, Architectures and Protocols Symposium. He chaired this Symposium, which continues to date under different names. He was the founding chair of the first Green Track at ICC/GLOBECOM at GLOBECOM 2011, and is Chair of the IEEE Sustainable ICT Initiative within the IEEE Technical Activities Board (TAB) Future Directions Committee (FDC) and within the IEEE Communications Society, a pan IEEE Societies Initiative responsible for Green and Sustainable ICT activities across IEEE, 2012-present. He is and has been on the technical program committee of 38 IEEE ICC/GLOBECOM conferences between 1995 and 2019 including 18 times as Symposium Chair. He received the IEEE Communications Society Hal Sobol award, the IEEE Comsoc Chapter Achievement award for excellence in chapter activities (both in 2005), the University of Wales Swansea Outstanding Research Achievement Award, 2006, the IEEE Communications Society Signal Processing and Communication Electronics outstanding service award, 2009, a best paper award at IEEE ICC'2013, the IEEE Comsoc Transmission Access and Optical Systems outstanding Service award 2015 in recognition of "Leadership and Contributions to the Area of Green Communications", received the GreenTouch 1000x award in 2015 for "pioneering research contributions to the field of energy efficiency in telecommunications", the 2016 IET Optoelectronics Premium Award and shared with 6 GreenTouch innovators the 2016 Edison Award in the "Collective Disruption" Category for their work on the GreenMeter, an international competition, clear evidence of his seminal contributions to Green Communications which have a lasting impact on the environment (green) and society. He is currently an editor of: IET Optoelectronics, Journal of Optical Communications, IEEE Communications Surveys and Tutorials and IEEE Journal on Selected Areas in Communications series on Green Communications and Networking. He was Co-Chair of the GreenTouch Wired, Core and Access Networks Working Group, an adviser to the Commonwealth Scholarship Commission, member of the Royal Society International Joint Projects Panel and member of the Engineering and Physical Sciences Research Council (EPSRC) College. He was Principal Investigator (PI) of the £6m EPSRC INTelligent Energy awaRe NETworks (INTERNET) Programme Grant, 2010-2016 and is currently PI of the £6.6m EPSRC Terabit Bidirectional Multi-user Optical Wireless System (TOWS) for 6G LiFi Programme Grant, 2019-2024. He has been awarded in excess of £30 million in grants to date from EPSRC, the EU and industry and has held prestigious fellowships funded by the Royal Society and by BT. He was an IEEE Comsoc Distinguished Lecturer 2013-2016.